# Effective Ways of Secure, Private and Trusted Cloud Computing

Pardeep Kumar[1], Vivek Kumar Sehgal[2], Durg Singh Chauhan[3], P. K. Gupta[4] and Manoj Diwakar[5]

[1] CSE & IT, Jaypee University of Information Technology
Waknaghat, Solan, Himachal Pradesh, 173215,India

[2] ECE, Jaypee University of Information Technology
Waknaghat, Solan, Himachal Pradesh , 173215,India

[3] Uttrakhand Technical University
Dehradun, Uttaranchal, 248007,India

[4] CSE & IT, Jaypee University of Information Technology
Waknaghat, Solan, Himachal Pradesh, 173215,India

[5] CSE & IT, Mody Institute of Technology & Science
Laxmangarh, Sikar, Rajasthan, 332311,India

## Abstract

Cloud computing is an Internet-based computing, where shared resources, software and information, are provided to computers and devices on-demand. It provides people the way to share distributed resources and services that belong to different organization. Since cloud computing uses distributed resources in open environment, thus it is important to provide the security and trust to share the data for developing cloud computing applications. In this paper we assess how can cloud providers earn their customers' trust and provide the security, privacy and reliability, when a third party is processing sensitive data in a remote machine located in various countries? A concept of utility cloud has been represented to provide the various services to the users. Emerging technologies can help address the challenges of Security, Privacy and Trust in cloud computing.

## 1. Introduction

Cloud computing is revolutionizing the way companies are implementing their information systems. Cloud computing has elevated IT to newer limits by offering the market environment data storage and capacity with flexible scalable computing processing power to match elastic demand and supply, whilst reducing capital expenditure [1]. As a result, Cloud adoption is spreading rapidly and represents a new opportunity that companies should not ignore given its profound impact.

Due to its perceived open nature, Cloud Computing raises strong security, privacy and trust concerns, namely:
- How data safely stored and handled by Cloud providers?
- How data privacy being managed adequately?
- Are Cloud providers adhering to laws and regulations?
- How is business disruption or outage kept to its minimum?
- Are Cloud providers sufficiently protected against cyber-attacks?

Cloud computing share distributed resources via the network in the open environment, thus it makes security problems important for us to develop the cloud computing application [2].Cloud computing means different things to different people. cloud computing is a convenient, on-demand model for network access to a shared pool of configurable computing resources (e.g., networks, servers, storage, applications, and services) that can be rapidly provisioned and released with minimal management effort or service provider interaction . The cloud element of cloud-computing derives from a metaphor used for the Internet, from the way it is often depicted in computer network diagrams. Conceptually it refers to a model of scalable, real-time, internet-based information technology services and resources, satisfying the computing needs of users, without the users incurring the costs of maintaining the underlying infrastructure. Examples in the private sector involve providing common business applications online, which are accessed from a web browser, with software and data stored on the "cloud" provider's servers.

Cloud computing provides many opportunities for enterprises by offering a range of computing services. It shares massively scalable, elastic resources (e.g., data, calculations, and services) transparently among the users over a massive network [3]. These opportunities, however, don't come without challenges. Cloud computing has opened up a new frontier of challenges by introducing a different type of trust scenario. Today, the problem of trusting cloud computing is a paramount concern for most enterprises. It's not that the enterprises don't trust the cloud providers' intentions; rather, they question cloud computing capabilities. Yet the challenges of trusting cloud computing don't lie entirely in the technology itself. The projected benefits of cloud computing are very compelling both from a cloud consumer as well as a cloud services provider perspective: ease of deployment of services; low capital expenses and constant operational expenses leading to variable pricing schemes and reduced opportunity costs; leveraging the economies of scale for both services providers and users of the cloud [4]. Unfortunately, the adoption of cloud computing came before the appropriate technologies appeared to tackle the accompanying challenges of trust. This gap between adoption and innovation is so wide that cloud computing consumers don't fully trust this new way of computing. To close this gap, we need to understand the trust issues associated with cloud computing from both a technology and business perspective. Then we'll be able to determine which emerging technologies could best address these issues.

## 2. Security challenges of Cloud computing

Cloud computing is not secure by nature. Security in the Cloud is often intangible and less visible, which inevitably creates a false sense of security and anxiety about what is actually secured and controlled. The off-premises computing paradigm that comes with cloud computing has incurred great concerns on the security of data, especially the integrity and confidentiality of data, as cloud service providers may have complete control on the computing infrastructure that underpins the services [5]. Accordingly, the various security challenges as shown in Fig-1, related to Cloud computing are worth of a deeper attention and can relate to many different aspects.

2.1 Information security policy –

Cloud users typically have no control over the Cloud resources used and there is an inherent risk of data exposure to third parties on the Cloud or the Cloud provider itself.

Some policies are

a) Whether there exists an Information security policy, which is approved by the management, published and communicated as appropriate to all employees.

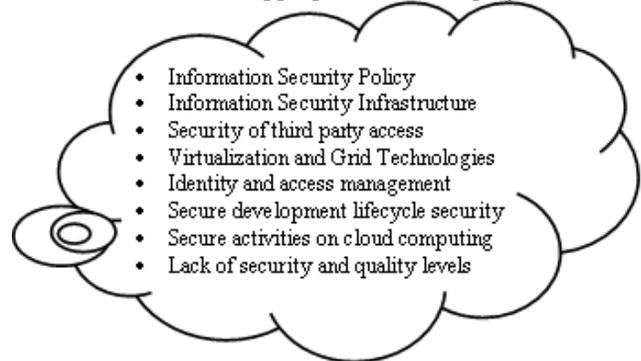

Fig. 1 Security Challenges of Cloud Computing

b) Whether it states the management commitment and set out the organizational approach to managing information security.
c) Whether the Security policy has an owner, who is responsible for its maintenance and review according to a defined review process.

Whether the process ensures that a review takes place in response to any changes affecting the basis of the original assessment, example: significant security incidents, new vulnerabilities or changes to organizational or technical infrastructure.

2.2. Information security infrastructure

The cloud computing architecture requires the adoption of identity and access management measures. Some issues:

a) Whether there is a management forum to ensure there is a clear direction and visible management support for security initiatives within the organization.
b) Whether there is a cross-functional forum of management representatives from relevant parts of the organization to coordinate the implementation of information security controls.
c) Whether responsibilities for the protection of individual assets and for carrying out specific security processes were clearly defined.
d) Whether there is a management authorisation process in place for any new information processing facility. This should include all new facilities such as hardware and software.
e) A specific individual may be identified to co-ordinate in-house knowledge and experiences to ensure

consistency, and provide help in security decision making.
f) Whether appropriate contacts with law enforcement authorities, regulatory bodies, information service providers and telecommunication operators were maintained to ensure that appropriate action can be quickly taken and advice obtained, in the event of a security incident.

Whether the implementation of security policy is reviewed independently on regular basis. This is to provide assurance that organizational practices properly reflect the policy, and that it is feasible and effective.

2.3. Security of third party access

To access the third party some challenges:
a) Whether risks from third party access are identified and appropriate security controls implemented.
b) Whether the types of accesses are identified, classified and reasons for access are justified.
c) Whether security risks with third party contractors working onsite was identified and appropriate controls are implemented.
d) Whether there is a formal contract containing, or referring to, all the security requirements to ensure compliance with the organization's security policies and standards.
e) Whether security requirements are addressed in the contract with the third party, when the organisation has outsourced the management and control of all or some of its information systems, networks and/ or desktop environments.

The contract should address how the legal requirements are to be met, how the security of the organization's assets are maintained and tested, and the right of audit, physical security issues and how the availability of the services is to be maintained in the event of disaster.

2.4. Virtualization and grid technologies

The virtual cloud essentially implements shared and coordinated task-spaces, which coordinates the scheduling of jobs submitted by a dynamic set of research groups to their local job queues [6]. Virtualization and grid technologies expose cloud infrastructures to emerging and high-impact threats against hypervisors and grid controllers.

2.5. Identity and access management

The cloud computing architecture requires the adoption of identity and access management measures. When data are trusted to a third party especially for handling or storage within a common user environment, appropriate precaution must be in place to ensure uninterrupted and full control of the data owner over its data.

2.6. Secure development lifecycle Security

Although traditional searchable encryption schemes allow users to securely search over encrypted data through keywords, these techniques support only Boolean search, without capturing any relevance of data files [7]. General purpose software, which was initially developed for internal use, is now being used within the cloud computing environment without addressing all the fundamental risks associated to this new technology. Another consequence of the migration to Cloud computing is that the secure development lifecycle of the organization may need to change to accommodate the Cloud computing risk context.

2.7. Secure activities on Cloud computing

As with most technological advances, regulators are typically in a "catch-up" mode to identify policy, governance, and law [8]. Migrating onto a Cloud may imply outsourcing some security activities to the Cloud provider. This may cause confusion between Cloud provider and user regarding individual responsibilities, accountability and redress for failure to meet required standards. Means to clarify those issues can be contracts, but also the adoption of policies, "service statements" or "Terms and Conditions" by the Cloud provider, which will clearly set forth obligations and responsibilities of all parties involved.

2.8 Lack of security and quality levels

Currently there is still a lack of generally-admissible Cloud computing standards at worldwide level. The consequence of this is uncertainty regarding the security and quality levels to be ensured by Cloud providers, but also vendor dependency for Cloud users given that every provider uses a proprietary set of access protocols and programming interfaces for their Cloud services.

## 3. Privacy challenges of Cloud computing

In the Cloud-computing environment, Cloud providers, being by definition third parties, can host or store important data, files and records of Cloud users. While outsourcing information to a cloud storage controlled by a cloud service provider relives an information owner of tackling instantaneous oversight and management needs, a significant issue of retaining the control of that information to the information owner still needs to be solved [9]. Given the volume or location of the Cloud computing providers, it is difficult for companies and private users to keep at all times in control the information or data they entrust to Cloud suppliers. Privacy is an important issue for cloud computing, both in terms of legal compliance and user trust, and needs to be considered at every phase of design [10]. Some key privacy or data protection challenges that can be characterized as particular to the Cloud-computing context are, in our view, as shown in fig-2 and described below:

### 3.1 Sensitivity of confidential information

Now, recession-hit companies are increasingly realizing that simply by tapping into the cloud they can gain fast access to best-of-breed business applications or drastically boost their infrastructure resources, all at negligible cost. But as more and more information on individuals and companies is placed in the cloud, concerns are beginning to grow about just how safe an environment [11]. It appears that any type of information can be hosted on, or managed by the Cloud. No doubt that all or some of this information may be business sensitive (i.e. bank account records) or legally sensitive (i.e. health records), highly confidential or extremely valuable as company asset (e.g. business secrets). Entrusting this information to a Cloud increases the risk of uncontrolled dissemination of that information to competitors (who can probably co-share same Cloud platform), individuals concerned by this information or to any other third party with an interest in this information.

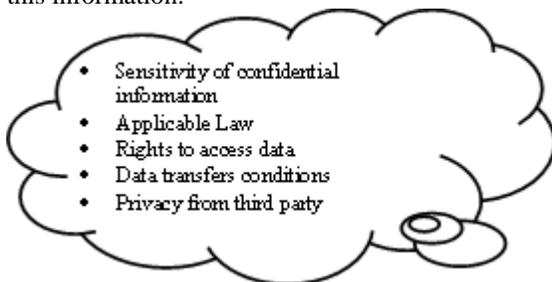

Fig. 2 Privacy challenges of cloud computing

### 3.2 Applicable law

Computer forensics is a relatively new discipline born out of the increasing use of computing and digital storage devices in criminal acts (both traditional and hi-tech). Computer forensic practices have been around for several decades and early applications of their use can be charted back to law enforcement and military investigations some 30 years ago [12].The relation of certain data to a geographic location has never been more blurred than with the advent of Cloud computing. As in other jurisdictions, the physical "location" plays a key role for determining which privacy rules apply. Thus, data collected and "located" within the European territory can benefit from the protection of the European privacy rules.

### 3.3 Rights to access data

Given that the users of the same Cloud share the premises of data processing and the data storage facilities, they are by nature exposed to the risk of information leakage, accidental or intentional disclosure of information.

### 3.4 Data transfers conditions

If the data used by, or hosted on, the Cloud may change location regularly or may reside on multiple locations at the same time, it becomes complicated to watch over the data flows and, consequently, to determine the conditions that would legitimize such data transfers. It may become complicated to fulfill these arrangements if data locations are not stable.

### 3.5 Privacy from third party

Companies engaging in Cloud computing expect that the privacy commitments they have made towards their customers, employees or other third parties will continue to apply by the Cloud computer provider. This becomes particularly relevant if the Cloud provider operates in many jurisdictions in which the exercise of individual rights may be subject to different conditions. Contractual rules with privacy implications It is common for a Cloud provider to offer his facilities to users without individual contracts. Yet, it can be that certain Cloud providers suggest negotiating their agreements with clients and, thus, offering the possibility of tailored contracts. Whatever the opted contractual model is, certain contractual clauses can have direct implications in the privacy and protection of the entrusted information (e.g. defining who actually "controls" the data and who only "processes" the data).

## 4. Trust challenges of Cloud computing

The Security and Privacy challenges discussed above are also relevant to the general requirement upon Cloud suppliers to provide trustworthy services. In the cloud computing, due to users directly use and operate the software and OS, and even basic programming environment and network infrastructure which provided by the cloud services providers, so the impact and destruction for the software and hardware cloud resources in cloud computing are worse than the current Internet users who use it to share resources [13].If Cloud providers find adequate solutions to address the data privacy and security specificities of their business model, they will have met in a certain way the requirement of offering trusted services. Yet, there are a few other challenges which, if tackled properly, would enhance user's confidence in the application of Cloud computing and would build market trust in the Cloud service offerings fig-3.

### 4.1 Trust in Infrastructure

The main difference between cloud computing and traditional enterprise internal IT services is that the owner and the user of cloud IT infrastructures are separated in cloud [14].One key to providing a trusted infrastructure is by provisioning a protected execution and content environment. Trusted Computing Group (TCG) and related open specifications and development efforts for servers, clients, and pervasive devices to provide a hardware "root of trust" that can leverage up the stack. Bindings that support end-to-end trust chains for web services and grid transactions.

TCG-enabled integrity reporting provides several capabilities for the trust framework to enable *trust in infrastructure*, including
a) Authentication of system configuration change origins.
b) Assertion of system platform identity and configuration.
c) Assertion of origin of execution image.
d) Verification of execution context.
e) Secure destruction of execution context.
f) IDS signature verification.
g) Signed, verifiable audit records.
h) Proof that audit logs were not tampered.
i) Validation of service provider.
j) Secure content management and distribution.

TCG support also provides a trust basis to support server cluster provisioning, to cover issues such as:
a) Do I let this new system into the cluster?
b) Does the system meet the trust requirements?
c) Does this system have the proper execution environment to interoperate in the cluster?
d) Has the system been tampered?

Is the system running as it was originally set up?

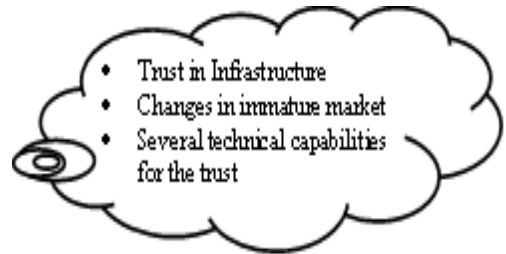

Fig. 3 Trust challenges of cloud computing

### 4.2 Changes in immature market

Cloud Computing Environments enable business agility and enterprises to exploit/respond quickly to changes in the marketplace, competition, technology and operational environment [15]. Accordingly, users of the Cloud must be confident that the services externalized to the Cloud provider, including any important assets.

### 4.3. Several technical capabilities for the trust

By definition, the Cloud-computing concept cannot guarantee full, continuous and complete control of the Cloud users over their assets. With public audit ability, a trusted entity with expertise and capabilities data owners do not possess can be delegated as an external audit party to assess the risk of outsourced data when needed [16]. Consider immediate investment in research and development activities to accelerate the maturation of the following commercial capabilities.
a) Development and acceptance of trust policy languages and trust management/negotiation protocols
b) Development and acceptance of trust inference engines, and definition of trust level semantics and assurance standards
c) Development and acceptance of privacy management technology
d) Development and acceptance of trusted identity management solutions that support federation (cross-domain entity resolution, credentialing, and access management).
e) Development and acceptance of secure development environments
f) Development and acceptance of key management/key exchange systems that can interoperate across trust domains and heterogeneous platforms

## 5. Enhanced Security, Privacy and Trust using cloud

It has the advantage of reducing cost by sharing computing and storage resources, combined with an on-demand provisioning mechanism relying on a pay-per-use business model. These new features have a direct impact on the budgeting of IT budgeting but also affect traditional security, trust and privacy mechanisms [17]. Besides the apparent challenges, Cloud computing can also lead to new opportunities in the fields of security, privacy and trust fig-4, for the Cloud users:
Effectively integrating security into a Manufacturing and Control System environment requires defining and executing a comprehensive program that addresses all aspects of security, ranging from identifying objectives to day-to-day operation and ongoing auditing for compliance and improvement.

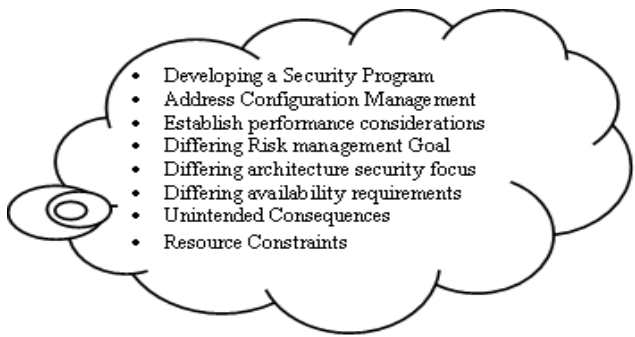

Fig. 4 Enhanced securities, privacy and trust

### 5.1 Address Configuration Management

Vital information and assets must be assessed and classified based on the consequences of loss, damage, or failure. Assign the appropriate levels of security protection and assess the vulnerability of Manufacturing and Control System Information loss or compromise.

### 5.2. Establish Performance Considerations

It is essential to review and consider the required performance at an overall systems level to ensure that they do not affect the required time-critical or other performance characteristics of these systems.

### 5.3 Differing risk management goals

Human safety and fault tolerance to prevent loss of life or endangerment of public health or confidence, loss of equipment, loss of intellectual property, or lost or damaged product.

### 5.4. Differing architecture security focus

In a typical IT system, the primary focus of security is protecting the information stored on the central server. In manufacturing systems, the situation is reversed. Edge clients are typically more important than the central server.

### 5.5. Differing availability requirements

Many manufacturing processes are continuous in nature. Unexpected outages of systems that control manufacturing processes are not acceptable. Exhaustive pre-deployment testing is essential to ensure high availability for the Manufacturing and Control System.

### 5.6. Unintended Consequences

All security functions integrated into the process control system must be tested to prove that they do not introduce unacceptable vulnerabilities. Adding any physical or logical component to the system may reduce reliability of the control system, but the resulting reliability should be kept to acceptable levels.

### 5.7. Resource constraints

Cloud computing can provide users dynamically scalable, shared resources over the internet, but users usually fear about security threats and loss of control of data and systems [18]. Manufacturing and Control Systems are resource constrained systems and do not include typical IT security technologies.

## 6. Designing and Implementing

The security program includes developing design models to describe the minimum acceptable recommended practices to be used in constructing a secure system as shown in fig-5. The suggested models:

### 6.1. Network Segments

The network comprising of a series of logical and physical layers can be divided into network segments to simplify the approach to designing secure network architecture. The network segments can be further classified as follows:
a) Enterprise Network Segment consisting of enterprise computer systems
b) Process Information Network Segment consisting of Manufacturing Execution System computers
c) Control network Segment consisting of controllers and Human Machine Interface devices
d) Field network Segment consisting of sensors and actuators
e) Process Segment consisting of pipes, valves, and transportation belts.

### 6.2. Access Control Model

This describes the recommended practices for accessing Manufacturing and Control Systems. This topic can be further sub-divided into the following topics:
a) User Access Management
b) User responsibilities
c) Network Access Control
d) Operating System Access Control
e) Application Access Control
f) Monitoring System Access and Use

The potential for security violations by providing greater control over the user's access to information and resources of multiple devices in a network. RBAC addresses the problem of traditional model by basing access on user's job responsibilities rather than customizing access for each individual. Hierarchical organization of the users to different levels and assigning group passwords reduces cost of maintenance and possible errors as the users' role.

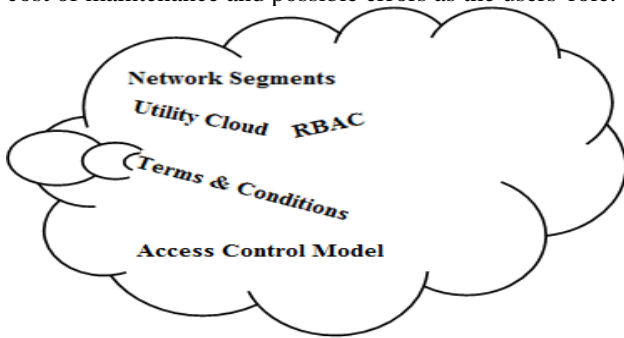

Fig. 5 Steps for designing and implementing a secure plan

RBAC tools offer interfaces to authorization mechanisms for most current platforms in the information technology area. However manufacturing and control systems are antiquated systems which do not support the new methods of RBAC. Also the centralized RBAC introduces additional points of failure that can impact the availability of the systems.

We apply to enable secure businesses through Cloud computing. Our risk based approach deals with these challenges:
a) How to control user from Cloud resources.
b) How to secure the private data.
c) How to use new technologies.
d) How to use and control the data.
e) How to provide security for confidential data.
f) How to maintain activities for Security models on Cloud computing.

Combining a sound integrated Security and Privacy Management with a clear view on the "to-be" architecture puts us in a unique position to define and execute a comprehensive security and privacy strategy. The design and implementation of security and privacy controls in an integrated manner cover a wide variety of methods that enable new business models while controlling the risks:

a) Determine if the strategy focuses on IT Disaster Recovery Planning, which may be limited to restoring cloud infrastructure at an alternative location or if it has a more holistic business orientation focusing on resuming all critical business operations.
b) Gather management's perception of the most critical business processes (and why), and management's formal/ informal assessment of the effectiveness of the company's ability to resume business operations in the event of a disruption. Determine if the business risks and impacts of unexpected disruptions have been identified and quantified by management.
c) Review the analysis of the organization's previous business continuity tests. Determine if the tests were successful. If not, why not? Identify recurring issues or other potential problem areas and understand the reasons for their existence.
d) Review documentation the organization has developed regarding business continuity processes, policies, standards and service level agreements. Determine if the processes are adequately documented, maintained and communicated to appropriate personnel.
e) Determine if a Business Continuity Plan exists and assess the degree to which it has been defined, documented, tested, maintained and communicated.
f) Determine the degree to which the organization uses software tools to facilitate Business Continuity Management processes.

Determine what Service Level Agreements are in place between the function chartered with Business Continuity Management, its various support organizations and other business units. Understand what the agreements include and discuss the criteria for determining SLA achievement.

### 6.4. Defining Utility Cloud-

The concept behind defining the utility cloud fig – 6 is to make such a cloud which is only designed for hosting various kinds of public services like video conferencing, accessing various TV channels, e-ticketing, various types of reservations etc. this cloud will consist the various services where a security and privacy is not required and people are free utilize the various cloud services at free of cost.

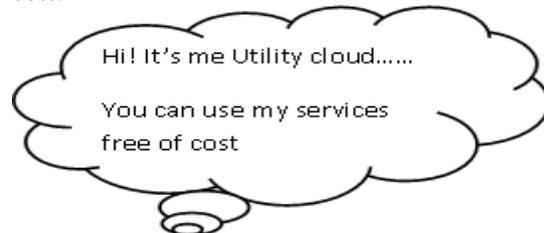

Fig. 6 Utility Cloud

## 6.5. Defining privacy and Terms & Conditions-

The regulatory landscape for Cloud computing is under continuous change and will still continue to evolve over the coming years. As such, contractual provisioning and Terms & Conditions are key components to seal trust, privacy and service levels in business relationships. The increasing availability of bandwidth allows new combinations and opens new IT perspectives [19]. Our team of world renowned lawyers specialized in Data Protection, Privacy, information technology law and outsourcing agreements develops pragmatic contractual templates that protect the business relationship. Additionally, we help government agencies and companies in data protection notifications for local data protection authorities regarding the collection and use of personal identifiable information.
We have extensive experience in dealing with such issues:
a) Sensitivity of entrusted information.
b) Localization of information and applicable law
c) User access rights to information
d) Cross border and third party data transfers
e) Externalization of privacy
f) Workable contractual rules with privacy implications

## 6.6 Managing risks and regulatory compliance-

Risk management framework is one of security assessment tool to reduction of threats and vulnerabilities and mitigates security risks [20].The risk management industry spans all other industries.  A quick internet search shows risk management associated with insurance, banking, financial services, IT, mining, environmental management, human resources, medicine, travel, forestry, asset management, energy, construction, sports, pharmaceuticals; basically risk management spans all of society.  Despite this wide-spread referencing, the most common association is between risk and insurance and/or financial services. Issues like:
a) Risks are effectively identified and evaluated.
b) Risk management processes are both effective and efficient.
c) Key risks are appropriately reviewed and reliably reported to those who need to know
d) While increasing attention is being paid to improving effectiveness, many companies are looking both to improve efficiencies and reduce the costs of effective governance, risk, and compliance activities.

## 6.7 Approach for quality assurance and conformity with requirements-

Board members and management seek assurance that their IT investment in Cloud computing does not jeopardize the confidential nature of critical business information. By providing approach Quality management systems, quality assurance and verification of conformity we tackle these challenges:
a) Determining the needs and expectations of customers and interested parties.
b) Establishing the quality policy and quality objectives of the organization.
c) Determining the processes and responsibilities necessary to attain the quality objectives.
d) determining and providing the resources
e) Establishing methods to measure the effectiveness and efficiency of each process.
f) Applying these measures to determine the effectiveness and efficiency of each process.
g) Determining means of preventing nonconformities and eliminating their causes.
h) Establishing and applying a process for continual improvement of the quality management system

## 7. CONCLUSIONS

Cloud providers need to safeguard the privacy and security of personal data that they hold on behalf of organizations and users [1]. Responsible management of personal data is a central part of creating the trust that underpins adoption of cloud based services – without trust, customers will be reluctant to use cloud-based services.
 We have analyzed the trusted computing in the cloud computing environment and the function of trusted computing platform in cloud computing. The advantages of our proposed approach are to extend the trusted computing technology into the cloud computing environment to achieve the trusted computing requirements for the cloud computing and then fulfill the trusted cloud computing.
The importance of trust varies from organization to organization, depending on the data's value. Furthermore, the less trust an enterprise has in the cloud provider, the more it wants to control its data—even the technology. However, it's crucial that consumers and providers change their mindsets. Trusting cloud computing might differ from trusting other systems, but the goal remains the same—improve business and remain competitive by exploiting the benefits of a new technology. Any new technology must gradually build its reputation for good performance and security, earning users' trust over time.
We will make more protocol to provide high security for Security management, Business continuity management, Identity & access management, Privacy & data protection and application Integrity in the future.

## References

[1] Ramgovind, S. Eloff and M.M. Smith, E.,"The management of security in Cloud computing", in Information Security for South Asia (ISSA), 2010, pp. 1-7.
[2] Zhidong Shen and Qiang Tong, "The security of cloud computing system enabled by trusted computing technology" in 2010 2nd International Conference on Signal Processing Systems" (ICSPS), 2010, pp. 2-11.
[3] Habib, S.M. Ries and S. Muhlhauser,"Cloud Computing Landscape and Research Challenges Regarding Trust and Reputation" in Ubiquitous Intelligence & Computing and 7th International Conference on Autonomic & Trusted Computing (UIC/ATC), Oct. 2010, pp. 410-444.
[4] Bhattacharya, Kamal Bichler, Martin Tai and Stefan, "ICSE Cloud 09: First international workshop on software engineering challenges for Cloud Computing" Software Engineering - Companion Volume, 2009.
[5] Gansen Zhao Chunming Rong Jin Li Feng Zhang Yong Tang, "Trusted Data Sharing over Untrusted Cloud Storage Providers" in 2010 IEEE Second International Conference on Cloud Computing Technology and Science (CloudCom), Nov 30 – Dec 3, 2010, pp. 97-103.
[6] Hyunjoo Kim, Parashar M., Foran D.J. and Lin Yang, "Investigating the use of autonomic cloudbursts for high-throughput medical image registration" in 2009 10th IEEE/ACM International Conference on Grid Computing (GRID 2009) , 2009.
[7] C. Wang, N. Cao, J. Li, K. Ren, and W. Lou, "Secure Ranked Keyword Search over Encrypted Cloud Data", in Proc. ICDCS, 2010, pp.253-262.
[8] Kaufman, L.M. , "Data Security in the World of Cloud Computing" in Security & Privacy, IEEE , July-Aug. 2009, Volume : 7 ,Issue:4 ,pp. 61-64.
[9] Ahmed, M., Xiang, Y., and Ali, S "Above the Trust and Security in Cloud Computing: A Notion Towards Innovation", in 2010 IEEE/IFIP 8th International Conference on Embedded and Ubiquitous Computing (EUC), Dec. 2010, pp. 723-730.
[10] Jian Wang ,Yan Zhao ,Shuo Jiang and Jiajin Le, "Providing privacy preserving in Cloud computing", in 2009 International Conference on Test and Measurement, 2009,pp. 213-216.
[11] Gansen Zhao Chunming Rong Jin Li Feng Zhang Yong Tang, "Trusted Data Sharing over Untrusted Cloud Storage Providers" in 2010 IEEE Second International Conference on Cloud Computing Technology and Science (CloudCom), Nov 30 – Dec 3, 2010, pp. 97-103.
[12] Reilly, D. Wren and C. Berry, "Cloud computing: Forensic challenges for law enforcement", in 2010 International Conference on Internet Technology and Secured Transactions (ICITST), Nov. 2010, pp. 1-7.
[13] Li-qin Tian , Chuang Lin and Yang Ni, "Evaluation of user behavior trust in cloud computing", in  2010 International Conference on Computer Application and System Modeling (ICCASM),Nov 4, 2010, pp. V7-567-V7-562.
[14] Xiao-Yong Li , Li-Tao Zhou ,Yong Shi and Yu Guo, "A trusted computing environment model in cloud architecture", in 2010 International Conference on Machine Learning and Cybernetics (ICMLC), July 2010, Volume 6, pp. 2843-2848.
[15] Goyal P,  "Enterprise Usability of Cloud Computing Environments: Issues and Challenges", in 2010 19th IEEE International Workshop on Enabling Technologies: Infrastructures for Collaborative Enterprises (WETICE), 2010,pp. 54-59.
[16] Cong Wang, Kui Ren ,Wenjing Lou and Jin Li, "Toward publicly auditable secure cloud data storage services", in IEEE Network, July-August 2010, Volume : 24 , Issue:4, pp.19-24.
[17] Pearson, S. and Azzedine Benameur, "Privacy, Security and Trust Issues Arising from Cloud Computing" in 2010 IEEE Second International Conference Cloud Computing Technology and Science (CloudCom),Nov 30-Dec 3,2010, page(s): 693-702.
[18] Jinzhu Kong, "A Practical Approach to Improve the Data Privacy of Virtual Machines" 2010 IEEE 10th International Conference on Computer and Information Technology (CIT), June 29 -July 1 ,2010, pp. 936-941.
[19] Esteves, R.M. and Chunming Rong, "Social Impact of Privacy in Cloud Computing" in 2010 IEEE Second International Conference on Cloud Computing Technology and Science (CloudCom), Nov. 30-Dec. 3 ,2010, pp. 593-596.
[20] Xuan Zhang Wuwong, N. Hao Li Xuejie Zhang , "Information Security Risk Management Framework for the Cloud Computing Environments" in 2010 IEEE 10[th] International Conference Computer and Information Technology (CIT), 2010,pp. 1328 – 1334.

**Pardeep Kumar** He has achieved his B.Tech degree from Kurukshetra University in 2004 and M.Tech degree from Guru Jambheshwar University of Science & Technology, Hisaar, Haryana in 2007. He has been associated with Mody Institute of Technology & Science, Sikar (Raj.)-India for one year. Currently he is working with Jaypee University of Information Technology, Solan (H.P)-India. He is a member of International Association of Engineers(IAENG) and its societies of Data Mining and Computer Science. His area of research is related to Knowledge Discovery in Databases(KDD) and Machine learning.

**Vivek Kumar Sehgal**  He has achieved his M.Tech degree from NSIT, Delhi and Ph.D degree from Uttrakhand Technical University, Dehradun . Currently he is working with Jaypee University of Information Technology, Solan, H.P-India. He has a membership of various technical bodies including  IEEE,ACM,IAENG .  His areas of interest include embedded processor architecture, hardware software co-design, smart sensors and soft computing for chip design.

**Durg Singh Chauhan** He did his B.Sc Engg.(1972) in electrical engineering at I.T. B.H.U., M.E. (1978) at R.E.C. Tiruchirapalli ( Madras University ) and Ph.D. (1986) at IIT/Delhi. His brilliant career brought him to teaching profession at Banaras Hindu University where he was Lecturer, Reader and then has been Professor till today. He has been director KNIT sultanpur in 1999-2000 and founder vice Chancellor of U.P.Tech. University (2000-2003-2006). Later on, he has served as Vice-Chancellor of Lovely Profession University (2006-07) and Jaypee University of Information Technology (2007-2009) Currently he has been serving as Vice-Chancellor of Uttarakhand Technical University.

**P.K. Gupta** graduated in Informatics and Computer Engineering from Vladimir State University, Vladimir, Russia, in 1999 and received his M.E. degree in Informatics & Computer Engineering in 2001 from the same university. He has been associated with academics more than ten years in different institutions like BIT M.Nagar, RKGIT Ghzaiabad. Currently he is working as Senior Lecturer with the Department of Computer Science and Engineering, Jaypee University of Information Technology, Waknaghat, Solan(HP), India. He is also pursuing his Ph.D. from JUIT Solan. He has supervised a number of B.Tech/M.Tech/M.Phil. theses from various universities of India. His research interests include Storage Networks, Green Computing, Software Testing, and Communication. P.K. Gupta is a Member of IEEE, Life Member of CSI and Life member of Indian Science Congress Association.

**Manoj Diwakar** He did his B.Tech degree from UPTU-Uttrar Pardesh Technical University and M.Tech from MITS, Gwalior. Currently he is working with Mody Institute of Technology & Science, Sikar (Raj.)-India. His areas of research include mobile and distributed computing.